\documentclass[aps,prl,twocolumn,superscriptaddress,showpacs,floatfix,nofootinbib]{revtex4}

\usepackage{graphicx}
\usepackage{epsfig}
\usepackage[english]{babel}

\newcommand{\be}{\begin{equation}}
\newcommand{\bea}{\begin{eqnarray}}
\newcommand{\eea}{\end{eqnarray}}
\newcommand{\ee}{\end{equation}}

\newcommand{\ket}[1]{\mbox{$| #1 \rangle$}}

\newcommand{\proj}[1]{\mbox{$|#1\rangle \!\langle #1 |$}}

\def\SU{{\cal SU}}
\def\U{{\cal U}}

\begin{document}

\title{A universal quantum circuit for two-qubit transformations \\with three CNOT gates}

\author{G. Vidal}
\affiliation{Institute for Quantum Information, California Institute of Technology, Pasadena, CA 91125 USA}
\author{C. M. Dawson}
\affiliation{Center for Quantum Computer Technology and Department of Physics, The University of Queensland, Brisbane 4072 Australia}

\date{\today}

\begin{abstract}
We consider the implementation of two-qubit unitary transformations by means of CNOT gates and single-qubit unitary gates. We show, by means of an explicit quantum circuit, that together with local gates three CNOT gates are necessary and sufficient in order to implement an arbitrary unitary transformation of two qubits. We also identify the subset of two-qubit gates that can be performed with only two CNOT gates.
\end{abstract}


\maketitle

In the context of establishing the existence of universal sets of two-qubit gates for quantum computation \cite{ll}, Barenco et al. \cite{Ba95} showed that any unitary transformation on $n$ qubits can be decomposed into a sequence of CNOT and single-qubit gates. Since then it has become customary to express $n$-qubit unitary transformations ---associated e.g. with quantum algorithms--- as a series of CNOT and single-qubit gates in the quantum circuit model \cite{NiCh}.

Relatedly, in those experimental settings where significant single-qubit control is already available, the ability to reliably perform a CNOT gate has become the standard hallmark of multi-qubit control. As a consequence, achieving a CNOT gate is one of the most popular and coveted goals among quantum information experimentalists \cite{Fort}. In turn, this has triggered theoretical studies on the optimal use of two-qubit interactions and of entangling gates to perform a CNOT gate \cite{Khaneja,VHC,Br,Zh}. 

In this paper we consider the construction of quantum circuits that minimize the use of CNOT gates. Such optimal constructions are of significant interest at least in two contexts.
First, they play a role in determining the algorithmic complexity of a given quantum computation, that is, the number of elementary gates required to implement the corresponding $n$-qubit unitary evolution. A most remarkable result of \cite{Ba95} is the explicit decomposition of an arbitrary $U \in \U(2^n)$ as a sequence of CNOT and single-qubit gates. This general construction, however, unavoidably requires exp($n$) CNOT gates, which renders the resulting quantum circuit inefficient. We recall, on the other hand, that the $n$-qubit unitary transformations relevant for quantum computation are precisely those that can be decomposed into only poly($n$) elementary gates for large $n$. Thus, given a unitary transformation $U\in \U(2^n)$ of interest, it is important to know how many CNOT gates are required to implement it.

Algorithmic complexity is typically concerned with gates involving a large number of qubits. But quantum circuits that minimize the use of CNOT gates are also of interest for gates involving only a reduced number of qubits, for a very practical reason: in present day experiments, two-qubit gates as the CNOT gate are imperfect due to technological limitations. Therefore, in order to minimize the probability that an error occurs in performing a certain unitary evolution $U\in\U(2^n)$, it is instrumental that the number of times the qubits interact is as small as possible. Unfortunately, it is not known in general how to optimally decompose $U$ into CNOT and single-qubit gates, not even for a small number of qubits, nor how to determine the minimal number of required CNOT gates. 

In this paper we describe a {\em universal} quantum circuit for two-qubit unitary transformations $U\in \SU(4)$ consisting of only three CNOT gates and four rounds of local gates. The shortest circuit previously known requires four CNOT gates \cite{BuM}. We also show that three CNOT gates are required in order to perform a generic two-qubit gate, thereby establishing the optimality of the proposed universal quantum circuit. Finally, we characterize the subset of two-qubit gates that require only two CNOT gates and construct an alternative, smaller quantum circuit for them.

We consider two qubits, labeled $A$ and $B$, and an arbitrary unitary transformation $U\in \SU(4)$. Let $u_l, v_l \in \SU(2)$ denote single-qubit unitary gates acting, respectively, on qubits $A$ and $B$, and let $U_{\mbox{\tiny CNOT}}$ denote a CNOT gate that has qubit $A$ as control and qubit $B$ as target, 
\be
U_{\mbox{\tiny CNOT}} ~ \ket{^zm}_A\otimes\ket{^zn}_B = \ket{^z m}_A\otimes\ket{^zn\oplus m}_B,~~~ m,n=0,1,
\label{eq:cnot}
\ee
where $i\oplus j$ denotes sum modulo 2 and $\ket{^z0}$ ($\ket{^z1}$) is the eigenvector with eigenvalue $1$ ($-1$) of the third of the Pauli matrices
\be
\sigma_x = \left( \begin{array}{cc}
0 & 1 \\
1 & 0
\end{array}\right),~~ 
\sigma_y = \left( \begin{array}{cc}
0 & -i \\
i & 0
\end{array}\right),~~ 
\sigma_z = \left( \begin{array}{cc} 
1 & 0 \\
0 & -1
\end{array}\right).~~ 
\label{eq:cnot2}
\ee

{\bf Theorem 1.---} {\em An arbitrary unitary gate
$U\in \SU(4)$
can be decomposed in terms of three CNOT gates and single-qubit unitary gates $u_l,v_l$ (to be specified below) as}
\be
\epsfig{file=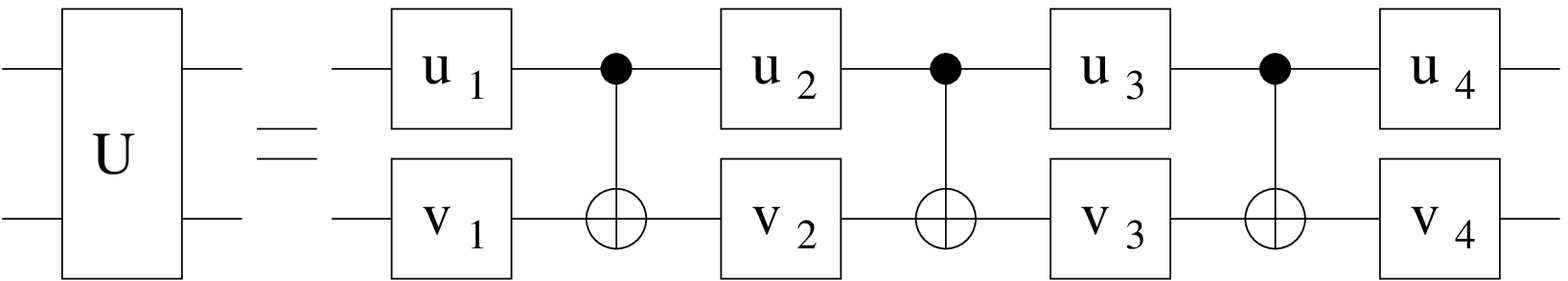,width=7cm}
\label{eq:deco}
\ee

An important element in order to prove theorem 1 is the decomposition of $U\in \SU(4)$ derived by Khaneja et al. \cite{Khaneja} and Kraus et al. \cite{Kraus}, namely
\bea
&&\epsfig{file=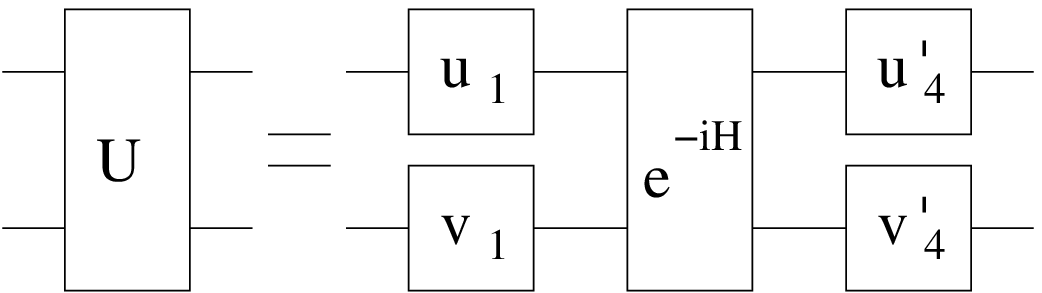,width=4.5cm} \label{eq:UH}\\
&&H \equiv h_x~ \sigma_x\otimes \sigma_x + h_y~\sigma_y\otimes \sigma_y + h_z~\sigma_z\otimes \sigma_z,
\label{eq:H}
\eea 
where $\pi/4 \geq h_x \geq h_y \geq |h_z|$. 
An explicit protocol to extract the single-qubit gates $u_1,v_1,u_4',v_4'\in\SU(2)$ and the coefficients $h_x, h_y, h_z \in {\cal R}$  from $U$ was presented in Ref. \cite{Kraus}. In what follows we show that $e^{-iH}$ can be further decomposed as
\be
\epsfig{file=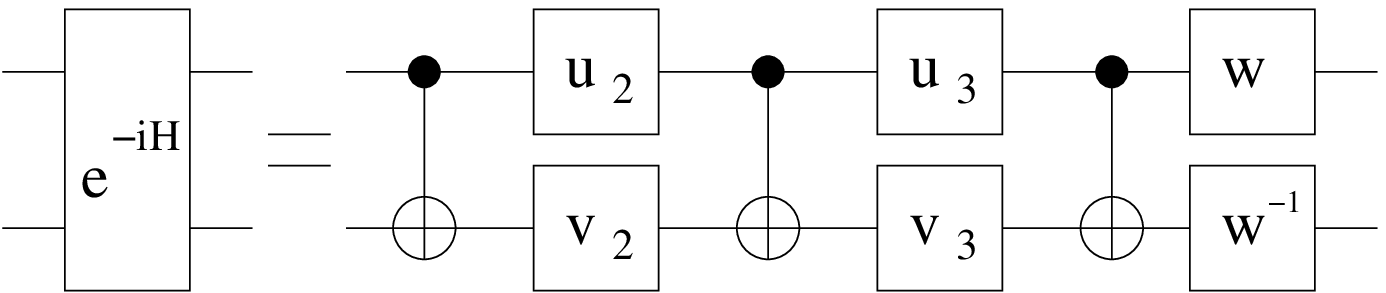,width=6cm}
\label{eq:Hexp}
\ee
where
\bea
u_2 \equiv \frac{i}{\sqrt{2}}(\sigma_x+\sigma_z)~e^{-i(h_x+\frac{\pi}{2})\sigma_x},&& v_2 \equiv  e^{-ih_z\sigma_z},\\
u_3 \equiv  \frac{-i}{\sqrt{2}}(\sigma_x+\sigma_z), ~~~~~~~~~~~~~~~~&&
v_3 \equiv  e^{ih_y\sigma_z},\\
w \equiv  \frac{I-i\sigma_x}{\sqrt{2}}, ~~~~~~~~~~~~~~~~~~~~~~~&&
\eea
so that $u_4$ and $v_4$ in (\ref{eq:deco}) are
\be
u_4 = u_4'w,~~~~~~~~~~~~~~~~~~~~v_4 = v_4'w^{-1}.
\ee

Let us introduce the Bell basis
\bea
\ket{\gamma_{00}} \equiv \frac{1}{\sqrt{2}} (\ket{00} + \ket{11}), &&
\ket{\gamma_{01}} \equiv \frac{1}{\sqrt{2}} (\ket{01} + \ket{10}),\nonumber\\
\ket{\gamma_{10}} \equiv \frac{1}{\sqrt{2}} (\ket{00} - \ket{11}), &&
\ket{\gamma_{11}} \equiv \frac{1}{\sqrt{2}} (\ket{01} - \ket{10}),\nonumber
\eea
where $\ket{mn}$ denotes $\ket{^zm}_A\otimes\ket{^zn}_B$. Operator $H$ in Eq. (\ref{eq:H}) can be rewritten as
\be
H = \sum_{m,n=0}^1 \lambda_{mn} \proj{\gamma_{mn}},
\ee 
with $\lambda_{mn}$ defined as
\bea
\lambda_{00} \equiv h_x - h_y + h_z,~~  &&
\lambda_{01} \equiv h_x + h_y - h_z,\\
\lambda_{10} \equiv -h_x + h_y + h_z, &&
\lambda_{11} \equiv -h_x - h_y - h_z.
\eea
Then $e^{-iH}$ becomes
\be
e^{-iH} = \sum_{m,n=0}^1 e^{-i\lambda_{mn}} \proj{\gamma_{mn}}.
\ee
Direct inspection shows that circuit (\ref{eq:Hexp}) indeed acts on the Bell basis $\ket{\gamma_{mn}}$ as \cite{phases}
\be
\ket{\gamma_{mn}} \longrightarrow e^{-i\lambda_{mn}}\ket{\gamma_{mn}},
\label{eq:ww}
\ee
thereby proving the theorem. We include some of the details. The first (leftmost) CNOT in (\ref{eq:Hexp}) maps the Bell basis into a product basis, namely
\be
\ket{\gamma_{mn}} \longrightarrow \ket{^xm}_A\otimes \ket{^zn}_B,
\ee
where $\ket{^x0} \equiv (\ket{^z0}+\ket{^z1})/\sqrt{2}$, $\ket{^x1} \equiv (\ket{^z0}-\ket{^z1})/\sqrt{2}$. The local transformation $u_2\otimes v_2$ introduces convenient phases $\phi_{mn}(h_x,h_y)$ into this local basis, and maps it into a new product basis,
\be
\ket{^xm}_A\otimes \ket{^zn}_B \longrightarrow e^{-i \phi_{mn}(h_x,h_z)} \ket{^zm}_A\otimes \ket{^zn}_B.
\ee
The second CNOT gate exchanges only two elements of the new product basis (recall Eq. (\ref{eq:cnot})),
\be
\ket{^z1}_A\otimes \ket{^z0}_B \leftrightarrow  \ket{^z1}_A\otimes \ket{^z1}_B,
\label{eq:perm}
\ee
after which $u_3\otimes v_3$ switches back to the $\ket{^x m}_A\otimes \ket{^z n}_B$ basis and introduces more phases $\phi_n'(h_y)$. The leftmost CNOT in (\ref{eq:Hexp}) maps the latter product basis back into the original Bell basis,
\be
\ket{^xm}_A\otimes \ket{^zn}_B \longrightarrow \ket{\gamma_{mn}},
\ee
and the final local gates $w\otimes w^{\dagger}$ exchange vectors $\ket{\gamma_{10}}$ and $\ket{\gamma_{11}}$ in order to undo the permutation (\ref{eq:perm})  [and also add a $\pi/4$ phase to each of them], so that circuit (\ref{eq:Hexp}) implements transformation (\ref{eq:ww}).

As shown in \cite{Ba95}, a nontrivial subset of two-qubit unitary transformations, namely control-V transformations for $V\in \U(2)$, can be performed by using only two CNOT gates and single-qubit gates. For these gates, one finds $h_y=h_z=0$ in its decomposition (\ref{eq:UH})-(\ref{eq:H}), so that they are locally equivalent to a control-phase gate $U_\varphi$,
\be
U_\varphi ~ \ket{^zm}_A\otimes\ket{^zn}_B = e^{-imn\varphi} \ket{^z m}_A\otimes\ket{^zn}_B,
\label{eq:cphi}
\ee
for an arbitrary phase $\varphi$. Theorem 2 characterizes the set of two-qubit transformations that can be performed with only two CNOT gates. They correspond to $h_z=0$ in (\ref{eq:UH})-(\ref{eq:H}), and are therefore a subset of zero measure in the space of two-qubit gates. 

{\bf Theorem 2.---} 
{\em A two-qubit gate $\bar{U}\in\SU(4)$ can be decomposed in terms of two CNOT gates and single-qubit gates $\bar{u}_l,\bar{v}_l$,  
\be
\epsfig{file=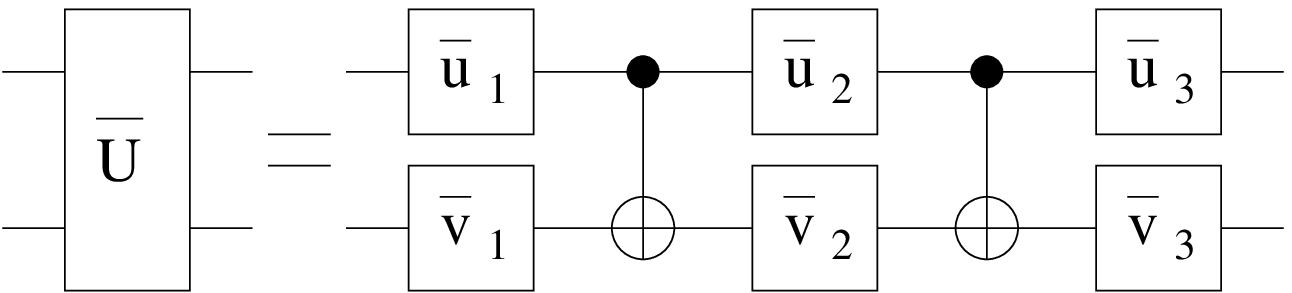,width=6cm}
\label{eq:2cnot}
\ee
if and only if $h_z=0$ in its decomposition (\ref{eq:UH})-(\ref{eq:H}).}

First we prove that if $h_z=0$, then $\bar{U}$ can be decomposed as in (\ref{eq:2cnot}). Decomposition (\ref{eq:UH})-(\ref{eq:H}) becomes
\bea
\epsfig{file=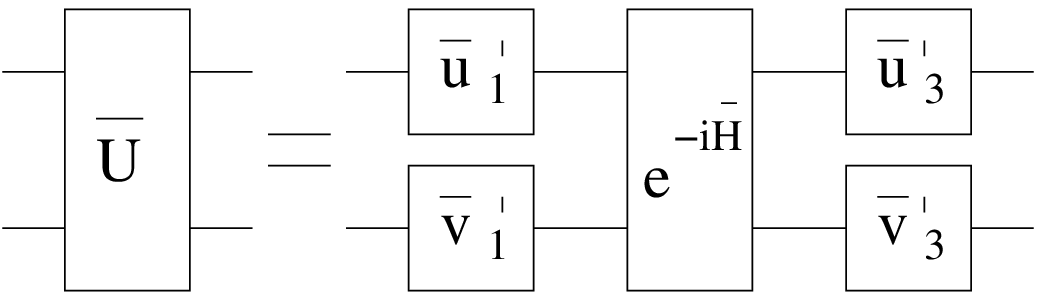,width=4.5cm}&&
\label{eq:UH2}\\
\bar{H} \equiv h_x~ \sigma_x\otimes \sigma_x + h_y~\sigma_y\otimes \sigma_y,&&h_x\geq h_y \geq 0.
\label{eq:H2}
\eea
One can now express $e^{-i\bar{H}}$ as 
\be
\epsfig{file=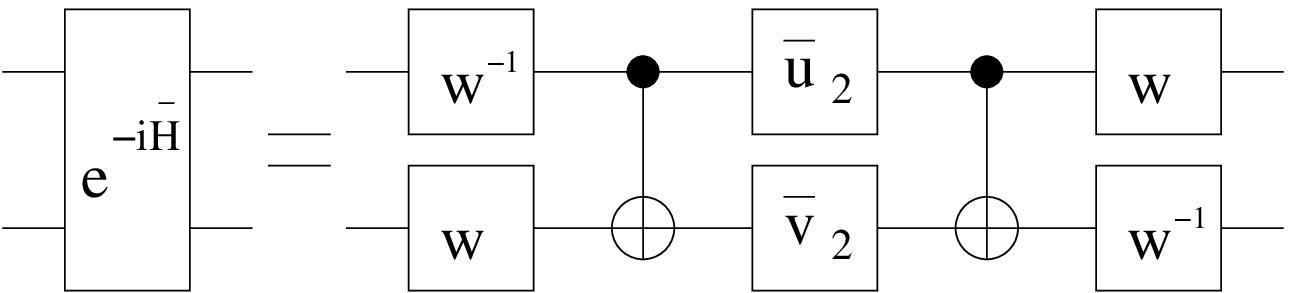,width=5cm}
\label{eq:eH2}
\ee
where
\be
\bar{u}_2 \equiv e^{-ih_x\sigma_x},~~  \bar{v}_2 \equiv  e^{-ih_y\sigma_z}, ~~
w \equiv  \frac{I-i\sigma_x}{\sqrt{2}}\ee
and $\bar{u}_l, \bar{u}'_l, \bar{v}_l, \bar{v}'_l$ in (\ref{eq:2cnot}) and (\ref{eq:UH2}) are related through
\be
\bar{u}_1 = w^\dagger\bar{u}_1',~~\bar{v}_1 = w\bar{v}_1',~~\bar{u}_3 = \bar{u}_3'w,~~\bar{v}_3 = \bar{v}_3'w^\dagger.
\ee
The validity of circuit (\ref{eq:eH2}) can be checked by reasoning similarly as we did in the proof of theorem 1. In particular, the first local gates $w^\dagger_A\otimes w_B$ permute vectors $\ket{\gamma_{10}}$ and $\ket{\gamma_{11}}$. Then the leftmost CNOT gate maps the Bell basis $\ket{\gamma_{mn}}$ into the product basis $\ket{^xm}_A\otimes\ket{^zn}_B$. Gates $\bar{u}_2\otimes\bar{v}_2$ introduce convenient phases $\phi_{mn}(h_x,h_y)$. Finally, the rightmost CNOT and $w_A\otimes w^{\dagger}_B$ gates map the local basis back into the original Bell basis.

In order to show that $h_z=0$ for any gate that can be decomposed in terms of only two CNOT gates and local gates, we assume the most general sequence of local gates and two CNOT gates. Up to initial and final local gates, this is equal to the quantum circuit on the LHS of (\ref{eq:commute}). We recall that, in much the same way as an arbitrary rotation $R\in SO(3)$ can be decomposed as three rotations along, say, axes $\hat{z}$, $\hat{x}$ and $\hat{z}$, gates $\bar{u},\bar{v} \in SU(2)$ can always be expanded as
\bea
\bar{u} = e^{-ia_u \sigma_z} ~\bar{u}_2(h_x) ~e^{-ic_u \sigma_z},\label{eq:u2}\\
\bar{v} = e^{-ia_v \sigma_x} ~\bar{v}_2(h_y) ~e^{-ic_v \sigma_x}.
\label{eq:v2}
\eea
for some value of $a_u,h_x,c_u,a_v,h_y,c_v \in {\cal R}$.
The leftmost and rightmost exponentials in Eqs. (\ref{eq:u2})-(\ref{eq:v2}) commute with the contiguous CNOT gates on the LHS of (\ref{eq:commute}), leading to the equivalence, up to local unitary transformations LU, expressed by
\be
\epsfig{file=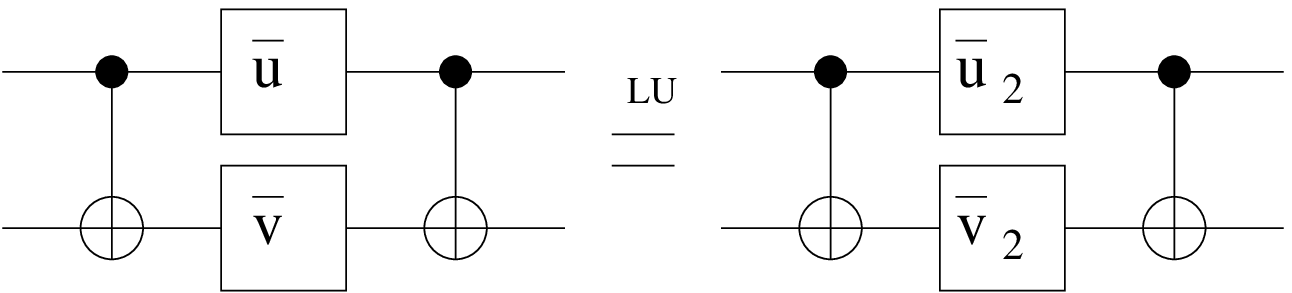,width=6cm}
\label{eq:commute}
\ee
But these circuits are also locally equivalent to that in (\ref{eq:eH2}), which corresponds to a gate $\bar{U}\in \SU(4)$ with $h_z=0$.

G.V. thanks J.I. Cirac and J. Pachos for comments and for hospitality at the Max Planck Institute for Quantum Optics, Garching, Germany, December 2001, and Debbie Leung for important corrections in an earlier version of this paper. This work was supported by the National Science Foundation of USA under grant EIA--0086038.


\begin{thebibliography}{99}
\bibitem{phases} In Eq. (\ref{eq:ww}) we have neglected a physically irrelevant, global phase $e^{i\pi/4}$ originating in that det$(U_{\mbox{\tiny CNOT}}) =-1$ in Eq. (\ref{eq:cnot}), i.e. $U_{\mbox{\tiny CNOT}}\notin \SU(4)$.
\end{thebibliography}
\end{document}